\documentclass[12pt]{article}
\usepackage[a4paper,margin=1in]{geometry}
\usepackage{bm}
\usepackage{amsmath}
\usepackage{amsfonts}
\usepackage{amssymb}
\usepackage{amsthm}
\theoremstyle{definition}
\newtheorem{theorem}{Theorem}
\newtheorem{proposition}{Proposition}
\newtheorem{lemma}{Lemma}
\newtheorem{corollary}{Corollary}
\newtheorem{remark}{Remark}
\newtheorem{example}{Example}
\usepackage{graphicx}
\usepackage{booktabs}
\usepackage{setspace}
\usepackage[natbibapa]{apacite}
\usepackage[font=small]{caption}
\usepackage[hidelinks]{hyperref}
\usepackage{tikz}
\usepackage{quantikz}
\newcommand{\Tr}{\operatorname{Tr}}
\title{Quantum Nondecimated Wavelet Transform: Theory, Circuits, and Applications}
\author{Brani Vidakovic\\Statistics Department, Texas A\&M University}
\date{November 2026\footnote{This is v2 version of the original December 2025 submission}}
\begin{document}
\maketitle

\begin{abstract}
The nondecimated wavelet transform (NDWT) is the translation-invariant member of the wavelet family and underlies cycle-spinning denoising and wavelet spectral estimation. We revisit the $\varepsilon$-decimated quantum NDWT of an earlier preprint, in which an $L$-qubit shift register prepared in uniform superposition controls circular shifts of an amplitude-encoded signal before a single orthogonal wavelet unitary of arbitrary compact support is applied. We prove that tracing out the shift register yields a reduced state whose level probabilities are exactly the normalized NDWT level energies, invariant to every circular shift, and whose diagonal is a moving average of squared NDWT coefficients. Computational-basis measurement gives a multinomial model for these level shares, and a single-qubit rotation gives the lag-one autocovariance of every detail level except the coarsest. These global, shift-invariant statistics are, however, estimated at least as precisely by a Fourier-basis measurement of the input state, which needs neither the shift register nor the wavelet unitary. The quantum NDWT is needed for localized statistics, for inexpensive marking oracles in amplitude estimation, and for coherent recombination of shift branches. For the last, we show that a block-encoded shrinkage operator placed between the transform and its inverse, with postselection of the shift register, implements the finite shift average of cycle spinning exactly as a linear combination of unitaries. Level-wise weights give a translation-equivariant zero-phase filter whose transfer function is a convex combination of wavelet band gains. Coordinatewise weights computed from the reduced state are not equivariant in general, and indexing them by the NDWT translate, through a block encoding controlled by the shift register, makes them equivariant. The same output maps can also be produced by direct NDWT analysis and synthesis or, for level-wise weights, by a Fourier-domain filter; what the shift-register circuit contributes is access to the individual decimated branches and their coherent recombination within one circuit. Weights estimated by measurement and feed-forward give a James--Stein-type level-wise estimator whose risk is close to that of the risk-optimal level-wise filter, and a translation-equivariant local empirical Wiener estimator. Gate-level simulations, a denoising study, and an illustrative scaling analysis of turbulent wind velocity accompany the theory, and the costs of state preparation, weight learning, and output are stated separately.

\vspace*{0.2in}

\end{abstract}

\section{Introduction}

The nondecimated wavelet transform (NDWT), known also as the stationary, maximal-overlap, or translation-invariant wavelet transform, removes the dyadic subsampling of the discrete wavelet transform (DWT) and returns, at every resolution level, one coefficient for every time point \citep{holschneider1989,shensa1992,NasonSilverman1995,percival2000}. The price is redundancy; the reward is that circular shifts of the input only shift the coefficient sequences. Two statistical uses of this property dominate practice. In nonparametric regression, cycle spinning averages a wavelet shrinkage estimator over shifts and removes the pseudo-Gibbs artifacts of decimated estimators \citep{CoifmanDonoho1995}. In spectral and scaling analysis, the average squared NDWT coefficient at each level is a shift-invariant estimator of the wavelet spectrum, whose log-linear decay identifies the Hurst exponent of self-similar processes \citep{Flandrin1992,AbryVeitch1998,percival2000}.

\citet{NasonSilverman1995} showed that the NDWT is equivalent to the collection of $2^L$ ordinary orthogonal DWTs of the circularly shifted signals, the $\varepsilon$-decimated transforms. Because each member of this collection is orthogonal, the collection is a natural object for quantum computation: the shift index can be promoted to a register, all shifted transforms can be prepared in coherent superposition, and a single application of the wavelet unitary acts on every branch. Orthogonal quantum wavelet transforms have been constructed by \citet{Hoyer1997TR}, \citet{FijanyWilliams1998LNCS}, \citet{Klappenecker1999}, and, for general wavelet families with polylogarithmic cost, by \citet{BagherimehrabAspuruGuzik2024}; extensions to multidimensional transforms, packets, and related multiscale transforms appear in \citet{LiEtAl2018QIP,LiEtAl2019QIP} and \citet{Podzorova2025}. \citet{nakahira2021} gave a direct unitary embedding of the nondecimated filter bank on $n+\lceil\log_2(L+1)\rceil$ qubits, without postselection. The shift-register construction of \citet{Vidakovic2025QNDWT}, studied here, is complementary to that embedding: it keeps the orthogonal transform intact and lets the redundancy reside in an auxiliary register of $L$ qubits, so that shift averaging is obtained by a partial trace and cycle-spun estimators are obtained by interference among shift branches.

This paper is a follow-up to the preprint \citet{Vidakovic2025QNDWT}, which introduced the coherent shift register, the partial trace over it, and level-projector readout. The present treatment adds the coverage lemma and exact normalization formulas for the reduced state, a sign-sensitive within-level observable, a comparison with Fourier-basis and randomized baselines that delimits what the shift register contributes, and a postselected circuit that reconstructs a cycle-spun output state, together with the conditions under which that output is translation equivariant. It also corrects two consequential claims of the preprint. The time-domain phase probe proposed there estimates a weighted local mass rather than a wavelet coefficient energy (Remark~\ref{rem:phase}), and the shrinkage described there as a trace-preserving channel cannot shrink an amplitude-encoded signal; shrinkage is the heralded, trace-decreasing branch of the circuit of Section~\ref{sec:shrink} (Remark~\ref{rem:channel}).

The aim of the present paper is to say precisely what statistical information this quantum object carries, how that information is extracted by measurement, and at what cost. Two facts frame the discussion. First, a quantum state stores the data only up to normalization, and the Born rule returns squared amplitudes; signed coefficients are not directly observable. Second, preparing the amplitude-encoded state of an arbitrary vector of length $N$ requires $\Theta(N)$ elementary gates in general \citep{shende2006,PleschBrukner2011,ZhangLiYuan2022}. A quantum NDWT therefore cannot accelerate the classical NDWT of a vector that is already stored in classical memory, and we do not claim that it does \citep{Aaronson2015}. Its use is as a subroutine on states produced coherently by an earlier quantum process, such as a simulation, a quantum sensor, or an efficient loader for structured data \citep{GroverRudolph2002}. Accordingly, all costs below are stated in two currencies: the number of elementary gates of the circuit acting on a given state, and the number of state preparations, or queries, needed to estimate a statistical quantity to a given precision.

Within this framework the paper makes four contributions. The first is structural. We prove that the reduced density operator of the data register has level probabilities equal to the normalized NDWT level energies, that these level shares are exactly invariant to all $N$ circular shifts of the input, and that the diagonal of the reduced state is a moving average of the squared NDWT coefficients. The second concerns measurement. Computational-basis measurement yields a multinomial model for the level shares with explicit variances, measuring the shift register as well identifies the energy of every individual NDWT coefficient, and a Hadamard gate on one data qubit yields the lag-one autocovariance of every detail level except the coarsest. We then show that the global statistics among these are estimated with no larger variance by measuring the input state in the Fourier basis, and that every statistic of the data register alone can be obtained by drawing the shift at random; the coherent shift register matters for amplitude estimation of the shares, whose marking oracle is a single multi-controlled $Z$ gate, and for recombination of the branches. The third contribution is an estimator. Cycle-spinning linear shrinkage over the finite set of $2^L$ shifts is implemented exactly by a linear combination of unitaries built from the quantum NDWT. With level-wise weights it is a translation-equivariant zero-phase filter; with coordinatewise weights it is equivariant when the weights are indexed by the NDWT translate, which we achieve with a block encoding controlled by the shift register, whereas index-wise weights computed from the reduced state are not equivariant in general. We also show that these output maps coincide with those of direct NDWT analysis and synthesis and, for level-wise weights, of a Fourier-domain filter, which locates the contribution of the shift register in the exposure and recombination of decimated branches rather than in the output map itself. Measurement and feed-forward of estimated energies give a James--Stein-type level-wise estimator and a translation-equivariant local empirical Wiener estimator. The fourth contribution is empirical: all constructions are verified at the gate level, the estimators are compared with the risk-optimal level-wise filter, with decimated and nonlinear benchmarks, on the test functions of \citet{donoho1994}, and the level-share spectrum is applied, as an illustration, to a record of turbulent wind velocity.

The paper is organized as follows. Section~\ref{sec:classical} fixes notation and records the combinatorial lemma on which the analysis rests. Section~\ref{sec:qed} constructs the quantum NDWT and characterizes its reduced state. Section~\ref{sec:measurement} develops measurement models and estimators and compares them with Fourier-basis and randomized baselines. Section~\ref{sec:shrink} treats translation-invariant shrinkage by linear combination of unitaries. Section~\ref{sec:wind} analyzes the wind data, and Section~\ref{sec:discussion} discusses resources, scope, and open problems. Python code reproducing every number and figure by statevector simulation, together with Qiskit notebooks \citep{Qiskit}, is available at {\tt github.com/BraniV/QNDWT}.

\section{Nondecimated Transform and $\varepsilon$-Decimation}
\label{sec:classical}

Let $x\in\mathbb{R}^N$ with $N=2^J$, and let $W$ be the $N\times N$ orthogonal matrix of a periodized DWT of depth $L\le J$ built from a compactly supported orthonormal wavelet with $2p$ filter taps \citep{Mallat1999,vidakovic1999}. We use Mallat's indexing, in which the data live in $V_J$ and
$$
V_J=V_{J-L}\oplus W_{J-L}\oplus W_{J-L+1}\oplus\cdots\oplus W_{J-1},
$$
so that $W_{J-1}$ is the finest detail space and $j$ decreases as resolution coarsens. The rows of $W$ are ordered accordingly. Row $i\in\{0,\ldots,2^{J-L}-1\}$ holds the scaling vector $\phi_{J-L,i}$, and row $i=2^j+k$, with $J-L\le j\le J-1$ and $0\le k<2^j$, holds the wavelet vector $g_{j,k}$. In binary, the level of a detail row is the position of the leading one of its index. The vectors at a level are circular translates of one another,
$$
g_{j,k}[n]=g_{j,0}[(n-2^{J-j}k)\bmod N],\qquad \phi_{J-L,k}[n]=\phi_{J-L,0}[(n-2^{L}k)\bmod N],
$$
so the decimated transform $w=Wx$ samples the level-$j$ filter output at stride $2^{J-j}$.

The NDWT retains every translate. Writing $g_{j,0}(\cdot-\tau)$ for the circular translate by $\tau\in\mathbb{Z}_N$, the NDWT of $x$ consists of the $L+1$ sequences
\begin{eqnarray}
d_j[\tau]=\langle g_{j,0}(\cdot-\tau),x\rangle,\qquad a[\tau]=\langle \phi_{J-L,0}(\cdot-\tau),x\rangle,\qquad \tau\in\mathbb{Z}_N,
\label{eq:ndwt}
\end{eqnarray}
each of length $N$, so that the NDWT has $(L+1)N$ coefficients. If $x$ is replaced by its circular shift, each sequence in (\ref{eq:ndwt}) is shifted by the same amount; this is the translation invariance that the decimated transform lacks.

Let $S$ denote the cyclic shift $S\ket{n}=\ket{(n+1)\bmod N}$, acting on vectors by $(Sx)[n]=x[(n-1)\bmod N]$. A change of summation index gives, for $i=2^j+k$,
\begin{eqnarray}
(WS^{\varepsilon}x)_{2^j+k}=d_j[2^{J-j}k-\varepsilon],\qquad (WS^{\varepsilon}x)_{k}=a[2^{L}k-\varepsilon],
\label{eq:branch}
\end{eqnarray}
with indices taken modulo $N$. Thus each shifted DWT $WS^\varepsilon x$ reads a decimated subsequence of every NDWT level, and the family $\{WS^\varepsilon x\}_{\varepsilon=0}^{2^L-1}$ is the $\varepsilon$-decimated library of \citet{NasonSilverman1995}. The following lemma records exactly how the library covers the NDWT; it is used repeatedly below.

\begin{lemma}[Coverage]\label{lem:coverage}
For $J-L\le j\le J-1$, the map $(\varepsilon,k)\mapsto (2^{J-j}k-\varepsilon)\bmod N$ from $\{0,\ldots,2^L-1\}\times\{0,\ldots,2^j-1\}$ onto $\mathbb{Z}_N$ is exactly $2^{L-J+j}$-to-one. The map $(\varepsilon,k)\mapsto(2^Lk-\varepsilon)\bmod N$ from $\{0,\ldots,2^L-1\}\times\{0,\ldots,2^{J-L}-1\}$ onto $\mathbb{Z}_N$ is a bijection. Moreover, for fixed $k$, the values $2^{J-j}k-\varepsilon$, $\varepsilon=0,\ldots,2^L-1$, are $2^L$ consecutive elements of $\mathbb{Z}_N$.
\end{lemma}

\begin{proof}
Write $\varepsilon=2^{J-j}u+v$ with $0\le v<2^{J-j}$ and $0\le u<2^{L-J+j}$. Then $2^{J-j}k-\varepsilon=2^{J-j}(k-u)-v$. Given $\tau\in\mathbb{Z}_N$, the residue $v\equiv-\tau \pmod{2^{J-j}}$ is determined, and $k-u$ is determined modulo $2^j$ by $(\tau+v)/2^{J-j}$. For each of the $2^{L-J+j}$ values of $u$ there is exactly one $k\in\{0,\ldots,2^j-1\}$. The scaling case is the case $j=J-L$ with $u=0$. The last statement is immediate.
\end{proof}

By Lemma~\ref{lem:coverage}, every NDWT coefficient of level $j$ appears exactly $2^{L-J+j}$ times in the library and every scaling coefficient exactly once. The NDWT is therefore a fixed linear selection from the library, and averaging over the $2^L$ library members any rule that applies the same function to every coefficient of a level is the same as averaging it over all $N$ circular shifts. This is the reason cycle spinning over $2^L$ shifts suffices for a depth-$L$ transform with level-dependent thresholding rules \citep{CoifmanDonoho1995}. A rule whose action depends on the position of a coefficient within its level does not have this property, a point that matters in Section~\ref{sec:shrink}.

\section{The Quantum $\varepsilon$-Decimated NDWT}
\label{sec:qed}

\subsection{Registers, Gates, and the Joint State}

The data register consists of $J$ qubits; qubit $q$ carries bit $b_q$ of the index $i=\sum_q b_q2^q$. The observed vector enters as the amplitude-encoded state
$$
\ket{y}=\sum_{n=0}^{N-1}y_n\ket{n},\qquad y=x/\|x\|,
$$
and the norm $\|x\|$, which the state cannot carry, is retained classically. We write $U_y$ for a unitary with $U_y\ket{0}=\ket{y}$. A shift register of $L$ qubits carries the index $\varepsilon=\sum_{l=0}^{L-1}\varepsilon_l2^l$. The controlled shift
$$
U_{\rm shift}=\sum_{\varepsilon=0}^{2^L-1}\ket{\varepsilon}\bra{\varepsilon}\otimes S^{\varepsilon}
$$
factors as $\prod_{l=0}^{L-1}(S^{2^l})^{\varepsilon_l}$, and $S^{2^l}$ is an increment of the $J-l$ most significant data qubits. Hence $U_{\rm shift}$ is a product of $L$ controlled increments, each realizable with $O(J)$ elementary gates and one ancilla qubit, or equivalently a single ripple-carry addition of the shift register into the data register \citep{Cuccaro2004}.

The wavelet unitary $W$ is applied once, to the data register only. For an arbitrary compactly supported orthonormal wavelet, each analysis step factors through the paraunitary lattice of the two-band polyphase matrix \citep{VaidyanathanHoang1988,Vaidyanathan1993},
$$
E(z)=\Theta_{p-1}\,\Lambda(z)\,\Theta_{p-2}\cdots\Lambda(z)\,\Theta_0,\qquad \Lambda(z)={\rm diag}(1,z^{-1}),
$$
where the $\Theta_r$ are plane rotations and $\Lambda(z)$ is a delay, the one-step circular shift of the odd samples. On a register of $m$ qubits, a rotation $\Theta_r$ is a single-qubit gate on the least significant qubit, and the delay is a decrement of the remaining $m-1$ qubits controlled by the least significant qubit, that is, a controlled permutation. A cyclic relabeling of the qubits then moves the high-pass outputs to the upper half of the index range, as Mallat's ordering requires, and the next level acts on the lower $m-1$ qubits. One level therefore costs $p$ single-qubit rotations and $p-1$ controlled decrements, and the full depth-$L$ transform costs $O(pLJ)$ elementary gates; polylogarithmic constructions of the same unitary are given by \citet{BagherimehrabAspuruGuzik2024}.

With the shift register prepared by $H^{\otimes L}$, the quantum NDWT is
\begin{eqnarray}
U_{\rm NDWT}=(I\otimes W)\,U_{\rm shift}=\sum_{\varepsilon=0}^{2^L-1}\ket{\varepsilon}\bra{\varepsilon}\otimes WS^{\varepsilon},
\label{eq:undwt}
\end{eqnarray}
and its action on the prepared input is
\begin{eqnarray}
\ket{\Psi}=U_{\rm NDWT}\big(H^{\otimes L}\ket{0^L}\otimes\ket{y}\big)=\frac{1}{2^{L/2}}\sum_{\varepsilon=0}^{2^L-1}\ket{\varepsilon}\otimes WS^{\varepsilon}\ket{y}.
\label{eq:psi}
\end{eqnarray}
The circuit is shown in Fig.~\ref{fig:arch}. Because $U_{\rm NDWT}$ is block diagonal with orthogonal blocks it is unitary; the shift register enlarges the Hilbert space to dimension $2^LN$ but does not multiply the number of wavelet transforms, since $W$ acts once on all branches. The construction uses $J+L$ qubits and $O(pLJ)$ elementary gates, exclusive of the preparation of $\ket{y}$. By (\ref{eq:branch}), the amplitude of $\ket{\varepsilon}\ket{2^j+k}$ in (\ref{eq:psi}) is $2^{-L/2}d_j[2^{J-j}k-\varepsilon]/\|x\|$, so by Lemma~\ref{lem:coverage} the state $\ket{\Psi}$ contains every NDWT coefficient, with a multiplicity that depends only on its level.

\begin{figure}[htb]
\centering
\begin{quantikz}[row sep=0.35cm, column sep=0.55cm]
\lstick{$\ket{0}_{\varepsilon_0}$} & \gate{H} & \ctrl{3} & \qw & \qw & \qw & \qw \\
\lstick{$\ket{0}_{\varepsilon_1}$} & \gate{H} & \qw & \ctrl{2} & \qw & \qw & \qw \\
\lstick{$\ket{0}_{\varepsilon_2}$} & \gate{H} & \qw & \qw & \ctrl{1} & \qw & \qw \\
\lstick{$\ket{y}$} & \qwbundle{J} & \gate{+1} & \gate{+2} & \gate{+4} & \gate{W} & \qw
\end{quantikz}
\caption{\small Quantum $\varepsilon$-decimated NDWT for $L=3$. The shift register is placed in uniform superposition; controlled increments by $2^l$ realize $S^\varepsilon$ on the data register; a single wavelet unitary $W$, built from paraunitary lattice rotations and delays, acts on all branches.}
\label{fig:arch}
\end{figure}

\subsection{The Reduced State and Exact Translation Invariance}

Most statistical uses of the NDWT are quadratic, and quadratic functionals of the data register are governed by the reduced density operator
$$
\rho_D=\Tr_S\ket{\Psi}\bra{\Psi}=\frac{1}{2^L}\sum_{\varepsilon=0}^{2^L-1}WS^{\varepsilon}\ket{y}\bra{y}S^{-\varepsilon}W^{T},
$$
obtained by discarding the shift register. The partial trace is a completely positive trace-preserving map, and $\rho_D$ is the uniform mixture of the $2^L$ decimated transforms. Let $P_j$ denote the projector onto the coordinates of level $j$, that is, $P_j=\ket{1}\bra{1}_{q_j}\otimes\bigotimes_{q>j}\ket{0}\bra{0}_{q}$, and let $P_s=\bigotimes_{q\ge J-L}\ket{0}\bra{0}_q$ be the projector onto $V_{J-L}$. Define the decimated level shares $p_j^{(\varepsilon)}=\|P_jWS^\varepsilon y\|^2$ and the quantum NDWT level shares $\bar p_j=\Tr(P_j\rho_D)$, and similarly $p_s^{(\varepsilon)}$ and $\bar p_s$.

\begin{theorem}\label{thm:ti}
For every $x\neq 0$ the following hold. (a) For $i=2^j+k$, the diagonal entry $\bra{i}\rho_D\ket{i}=2^{-L}\sum_{\varepsilon=0}^{2^L-1}d_j[2^{J-j}k-\varepsilon]^2/\|x\|^2$ is the average of the squared NDWT coefficients of level $j$ over a window of $2^L$ consecutive positions. (b) The level shares are normalized NDWT energies,
\begin{eqnarray}
\bar p_j=\frac{2^{j-J}}{\|x\|^2}\sum_{\tau\in\mathbb{Z}_N}d_j[\tau]^2,\qquad \bar p_s=\frac{2^{-L}}{\|x\|^2}\sum_{\tau\in\mathbb{Z}_N}a[\tau]^2,
\label{eq:shares}
\end{eqnarray}
and $\bar p_s+\sum_{j=J-L}^{J-1}\bar p_j=1$. (c) The level shares are invariant to all circular shifts: $\bar p_j(S^mx)=\bar p_j(x)$ and $\bar p_s(S^mx)=\bar p_s(x)$ for every $m\in\mathbb{Z}_N$. The decimated shares $p_j^{(0)}(S^mx)$ depend on $m$ through $m\bmod 2^{J-j}$ and are in general not constant in $m$. (d) The purity of the reduced state is
$$
\Tr\rho_D^2=\frac{1}{2^{2L}}\sum_{|\delta|<2^L}(2^L-|\delta|)\,r(\delta)^2,\qquad r(\delta)=\sum_{n}y_ny_{(n+\delta)\bmod N},
$$
a Bartlett-weighted sum of squared circular autocorrelations of the normalized data.
\end{theorem}

\begin{proof}
Part (a) follows from (\ref{eq:branch}) and the last statement of Lemma~\ref{lem:coverage}. For (b), summing (a) over $k$ and using the $2^{L-J+j}$-to-one coverage gives $\bar p_j=2^{-L}2^{L-J+j}\sum_\tau d_j[\tau]^2/\|x\|^2$; the scaling case uses the bijection, and the shares sum to one because $\Tr\rho_D=1$. For (c), the NDWT sequences of $S^mx$ are the sequences of $x$ shifted by $m$, and the sums in (\ref{eq:shares}) run over all of $\mathbb{Z}_N$; the norm is also shift invariant. For the decimated shares, (\ref{eq:branch}) shows that $p_j^{(0)}(S^mx)$ sums $d_j^2$ over the coset $-m+2^{J-j}\mathbb{Z}$, which depends on $m\bmod 2^{J-j}$. For (d), $\Tr\rho_D^2=2^{-2L}\sum_{\varepsilon,\varepsilon'}\langle y,S^{\varepsilon'-\varepsilon}y\rangle^2$ because $W$ is orthogonal, and the number of pairs with $\varepsilon'-\varepsilon=\delta$ is $2^L-|\delta|$.
\end{proof}

Theorem~\ref{thm:ti} is the quantum counterpart of the selection that maps the $\varepsilon$-decimated library to the NDWT. That selection is linear but not unitary and cannot be applied to a state; its role for quadratic summaries is taken by the partial trace, a physically realizable operation that requires no gates at all, since the shift register is simply not measured. The invariance in part (c) is exact and holds for all $N$ shifts, not only for the $2^L$ shifts present in the register. Part (d) shows that the mixedness of $\rho_D$ has a statistical reading: a signal with slowly decaying autocorrelation yields a nearly pure reduced state, because its shifted copies are nearly parallel, while white noise yields purity close to $2^{-L}$. The purity is estimable by a swap test on two copies of $\rho_D$ \citep{BuhrmanCleveWatrousDeWolf2001}.

\begin{example}[Haar transform of length eight]\label{ex:haar}
Let $J=3$, $L=2$, Haar filters, and $y=(5,3,19,1,7,-19,5,9)/\sqrt{912}$. The data space decomposes as $V_3=V_1\oplus W_1\oplus W_2$, with indices $\{0,1\}$, $\{2,3\}$, and $\{4,\ldots,7\}$. Table~\ref{tab:haar} lists the four branches $WS^\varepsilon y$ of (\ref{eq:psi}), whose amplitudes in $\ket{\Psi}$ carry the additional factor $1/2$, together with their level shares. The finest-level coefficients of the branches $\varepsilon$ and $\varepsilon+2$ are cyclic permutations of each other, as (\ref{eq:branch}) predicts, so $p_2^{(\varepsilon)}$ depends only on $\varepsilon\bmod 2$, while $p_1^{(\varepsilon)}$ and $p_s^{(\varepsilon)}$ take four different values. The quantum NDWT shares are the column averages $(\bar p_s,\bar p_1,\bar p_2)=(0.2445,0.2336,0.5219)$, which are the same for all eight circular shifts of $y$.
\end{example}

\begin{table}[htb]
\centering
{\scriptsize\setlength{\tabcolsep}{3.5pt}
\begin{tabular}{c|rr|rr|rrrr|ccc}
\toprule
$\varepsilon$ & $s_{1,0}$ & $s_{1,1}$ & $w_{1,0}$ & $w_{1,1}$ & $w_{2,0}$ & $w_{2,1}$ & $w_{2,2}$ & $w_{2,3}$ & $p_s^{(\varepsilon)}$ & $p_1^{(\varepsilon)}$ & $p_2^{(\varepsilon)}$\\
\midrule
0 & 0.4636 & 0.0331 & $-0.1987$ & $-0.4305$ & 0.0468 & 0.4215 & 0.6088 & $-0.0937$ & 0.2160 & 0.2248 & 0.5592\\
1 & 0.5960 & $-0.0993$ & $-0.1325$ & 0.3642 & 0.0937 & $-0.3746$ & $-0.1405$ & $-0.5620$ & 0.3651 & 0.1502 & 0.4846\\
2 & 0.3642 & 0.1325 & 0.0993 & 0.5298 & $-0.0937$ & 0.0468 & 0.4215 & 0.6088 & 0.1502 & 0.2906 & 0.5592\\
3 & 0.0000 & 0.4967 & $-0.4636$ & 0.2318 & $-0.5620$ & 0.0937 & $-0.3746$ & $-0.1405$ & 0.2467 & 0.2686 & 0.4846\\
\midrule
\multicolumn{9}{r|}{quantum NDWT shares $\bar p$} & 0.2445 & 0.2336 & 0.5219\\
\bottomrule
\end{tabular}
}
\caption{\small Branches $WS^\varepsilon y$ of the Haar quantum NDWT for $N=8$ and $L=2$ in Mallat order, with decimated level shares and their average, the reduced-state level shares $\bar p$.}
\label{tab:haar}
\end{table}

\begin{figure}[htb]
\centering
\includegraphics[width=\linewidth]{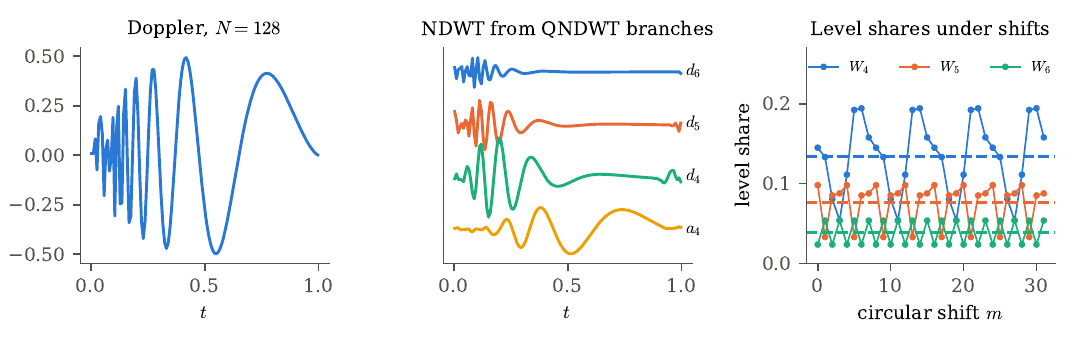}
\caption{\small Doppler signal of length $N=128$ (left); NDWT sequences $d_6,d_5,d_4$ and $a_4$ read from the amplitudes of the db2 quantum NDWT state with $L=3$ (middle); decimated level shares $p_j^{(0)}(S^mx)$ as functions of the circular shift $m$ (solid, shown for $m<32$) and the shift-invariant quantum NDWT shares $\bar p_j$ (dashed) (right).}
\label{fig:doppler}
\end{figure}

\begin{example}[Doppler signal with an arbitrary wavelet]\label{ex:doppler}
Let $N=128$ ($J=7$), $L=3$, and let $W$ be the Daubechies transform with two vanishing moments (db2) built from the lattice factorization above, so that $V_7=V_4\oplus W_4\oplus W_5\oplus W_6$. The statevector of $\ket{\Psi}$ was computed gate by gate, and the NDWT sequences were read from its amplitudes through (\ref{eq:branch}). The maximal discrepancy from the classical NDWT was $3.9\times10^{-16}$ (Fig.~\ref{fig:doppler}, middle). The right panel of Fig.~\ref{fig:doppler} shows the decimated shares $p_j^{(0)}(S^mx)$ as functions of the circular shift $m$. The share of $W_4$ ranges from $0.054$ to $0.194$, a factor of 3.6, depending on an arbitrary choice of origin, whereas the quantum NDWT share equals $0.1335$ for all $128$ shifts to within $2\times10^{-16}$.
\end{example}

\section{Measurement Models and Estimation}
\label{sec:measurement}

\subsection{Measurement in the Computational Basis}

Measuring the data register of $\rho_D$ in the computational basis returns an index $i$ with probability $\bra{i}\rho_D\ket{i}$, and the level label $\ell(i)$, the position of the leading one of $i$ or $s$ if $i<2^{J-L}$, is a deterministic function of the outcome. Since repeated preparations are independent, the following is immediate.

\begin{proposition}\label{prop:multinomial}
Let $n$ preparations of $\ket{\Psi}$ be measured in the computational basis on the data register, and let $n_j$ and $n_s$ be the numbers of outcomes with level labels $j$ and $s$. Then $(n_s,n_{J-L},\ldots,n_{J-1})\sim{\rm Multinomial}(n;\bar p_s,\bar p_{J-L},\ldots,\bar p_{J-1})$. The estimators $\hat p_j=n_j/n$ are unbiased and are the maximum likelihood estimators, with ${\rm Var}(\hat p_j)=\bar p_j(1-\bar p_j)/n$ and ${\rm Cov}(\hat p_j,\hat p_k)=-\bar p_j\bar p_k/n$. By the delta method, ${\rm Var}(\log_2\hat p_j)\approx(1-\bar p_j)/(n\,\bar p_j\ln^22)$.
\end{proposition}

Three practical consequences follow. First, a single batch of measurements estimates all $L+1$ shares simultaneously, with the covariance structure of the multinomial, which a generalized least-squares fit of a log-linear spectrum should respect. Second, the relative precision of $\hat p_j$ is governed by $n\bar p_j$, the expected number of counts at level $j$. Because the amplitude encoding normalizes by the total energy, removing the sample mean classically before loading moves energy out of $V_{J-L}$ and can reduce the required number of preparations by orders of magnitude; for the wind data of Section~\ref{sec:wind}, centering raises the total detail share from $0.0067$ to $0.57$, an 85-fold saving for the same relative precision. Third, if the shift register is measured as well, the joint outcome $(\varepsilon,\ell(i))$ has probability $2^{-L}p_{\ell}^{(\varepsilon)}$, so the same data yield the decimated shares of every branch. The $2^L\times(L+1)$ contingency table of these counts supports a Pearson chi-squared test of the hypothesis that the decimated shares do not depend on the shift, a direct test of whether a decimated analysis would be sensitive to the choice of origin. Finer still, by (\ref{eq:branch}) the outcome $(\varepsilon,i)$ with $i=2^j+k$ has probability $2^{-L}d_j[2^{J-j}k-\varepsilon]^2/\|x\|^2$, and aggregating outcomes by the translate $\tau=2^{J-j}k-\varepsilon$, which by Lemma~\ref{lem:coverage} is reached by $2^{L-J+j}$ pairs, gives
\begin{eqnarray}
\Pr(\text{level }j,\ \text{translate }\tau)=\frac{2^{j-J}d_j[\tau]^2}{\|x\|^2},\qquad \Pr(\text{scaling},\ \tau)=\frac{2^{-L}a[\tau]^2}{\|x\|^2}.
\label{eq:translate}
\end{eqnarray}
The joint record therefore estimates the energy of every individual NDWT coefficient, a localized statistic used in Section~\ref{sec:shrink}.

When a level receives no counts, $\log\hat p_j$ is undefined. In the simulations below a zero count is replaced by one half, a continuity correction whose effect is negligible at the budgets used; at the budgets of Figure~\ref{fig:readout} and Section~\ref{sec:wind} no zero counts occurred.

Absolute energies require the classical norm: the NDWT energy of level $j$ is $\sum_\tau d_j[\tau]^2=2^{J-j}\|x\|^2\bar p_j$. Ratios of shares, and therefore slopes of log-share spectra, do not.

\subsection{Sign-Sensitive Statistics by Basis Rotation}

Because the Born rule squares amplitudes, the computational basis reveals energies but not the signs of coefficients. Signs enter through observables that are not diagonal in that basis. For a Hermitian $O$ on the data register,
$$
\Tr(O\rho_D)=\frac{1}{2^L}\sum_{\varepsilon=0}^{2^L-1}\langle WS^\varepsilon y,\,O\,WS^\varepsilon y\rangle,
$$
so every such observable defines a shift-averaged quadratic form in the decimated coefficients. A particularly simple choice exposes the second-order dependence of wavelet coefficients within a level.

\begin{proposition}\label{prop:lag}
Let $X_0$ be the Pauli $X$ operator on data qubit $0$. For $\max(1,J-L+1)\le j\le J-1$, $X_0$ commutes with $P_j$ and
\begin{eqnarray}
\Tr(X_0P_j\rho_D)=\frac{2^{j-J}}{\|x\|^2}\sum_{\tau\in\mathbb{Z}_N}d_j[\tau]\,d_j[\tau+2^{J-j}],
\label{eq:lag}
\end{eqnarray}
the lag-$2^{J-j}$ circular autocovariance of the NDWT sequence of level $j$, that is, the covariance between neighboring coefficients in the units of that level. The quantity is estimated by applying a Hadamard gate to qubit $0$ immediately before computational-basis measurement and averaging $(-1)^{b_0}$ over the outcomes with level label $j$. The ratio $\hat r_j=\sum(-1)^{b_0}\mathbf{1}\{\ell=j\}/n_j$, defined when $n_j>0$, estimates the lag-one autocorrelation $r_j=\Tr(X_0P_j\rho_D)/\bar p_j$ with approximate variance $(1-r_j^2)/(n\bar p_j)$. The lag is $2^{J-j}$ samples on the NDWT grid, one step on the decimated grid of level $j$, and the coverage condition $2^{J-j+1}\le2^L$ excludes the coarsest detail level $j=J-L$.
\end{proposition}

\begin{proof}
For $j\ge1$ the operator $X_0$ flips the least significant bit and leaves the leading one at position $j$ unchanged, so it maps the level-$j$ subspace into itself and commutes with $P_j$. It pairs the indices $2^j+k$ and $2^j+k+1$ for even $k$, so $\Tr(X_0P_j\rho_D)=2^{-L}\sum_\varepsilon\sum_{k\,{\rm even}}2c^{\varepsilon}_{2^j+k}c^{\varepsilon}_{2^j+k+1}$ with $c^\varepsilon=WS^\varepsilon y$. By (\ref{eq:branch}) each product equals $d_j[\tau]d_j[\tau+2^{J-j}]/\|x\|^2$ with $\tau=2^{J-j}k-\varepsilon$. For even $k$, $\tau\equiv-\varepsilon\pmod{2^{J-j+1}}$, and when $2^{J-j+1}\le 2^L$ the argument of Lemma~\ref{lem:coverage} with stride $2^{J-j+1}$ shows that every $\tau\in\mathbb{Z}_N$ occurs exactly $2^{L-J+j-1}$ times. Collecting constants gives (\ref{eq:lag}). Conditional on $\ell=j$, the sign $(-1)^{b_0}$ after the Hadamard gate is a $\pm1$ variable with mean $r_j$, which gives the variance.
\end{proof}

The rotation does not disturb the level label, so a single measurement record yields both the shares of Proposition~\ref{prop:multinomial} and the autocorrelations of Proposition~\ref{prop:lag}. For the coarsest level $j=J-L$ the same measurement gives only the average over even pairings, which is not a circular autocovariance of the NDWT sequence. Other Pauli strings give other shift-averaged quadratic forms, and randomized measurements estimate many of them from a common record \citep{HuangKuengPreskill2020}.

\subsection{Fourier-Basis and Randomized Baselines}

The statistics of Proposition~\ref{prop:multinomial} and Proposition~\ref{prop:lag} are quadratic forms in $y$ that do not change when $y$ is shifted. Any such form is diagonal in the Fourier basis, which suggests a simpler measurement. Let $\hat y=F\ket{y}$, where $F$ is the unitary discrete Fourier transform, and let $A_j(\omega)$ and $A_s(\omega)$ be the band gains of Corollary~\ref{cor:filter} in Section~\ref{sec:shrink}, whose proof uses only Lemma~\ref{lem:coverage}.

\begin{proposition}\label{prop:qft}
Let $m$ be the outcome of a computational-basis measurement of $F\ket{y}$, so that $\Pr(m)=|\hat y_m|^2$, and let $\omega_m=2\pi m/N$. Then $A_j(\omega_m)$ is an unbiased estimator of $\bar p_j$ with per-shot variance $\sum_mA_j(\omega_m)^2|\hat y_m|^2-\bar p_j^2\le\bar p_j(1-\bar p_j)$, and likewise for $A_s$; a single outcome estimates all levels simultaneously. For $\max(1,J-L+1)\le j\le J-1$, $A_j(\omega_m)\cos(2^{J-j}\omega_m)$ is an unbiased estimator of $\Tr(X_0P_j\rho_D)$ with per-shot variance at most $\bar p_j-\Tr(X_0P_j\rho_D)^2$, the per-shot variance of the estimator of Proposition~\ref{prop:lag}.
\end{proposition}

\begin{proof}
By Theorem~\ref{thm:ti} and the definition of $T_j$ in Corollary~\ref{cor:filter}, $\bar p_j=\langle y,T_jy\rangle$, and $T_j$ is circulant with eigenvalues $A_j(\omega_m)$, so $\bar p_j=\sum_mA_j(\omega_m)|\hat y_m|^2$. Since $0\le A_j\le1$, $A_j^2\le A_j$ and the variance bound follows. By (\ref{eq:lag}), $\Tr(X_0P_j\rho_D)=\langle y,C_jy\rangle$ with $C_j=2^{j-J}\sum_\tau\frac12\{g_{j,\tau}g_{j,\tau+2^{J-j}}^T+g_{j,\tau+2^{J-j}}g_{j,\tau}^T\}$, writing $g_{j,\tau}$ for the translate $g_{j,0}(\cdot-\tau)$; $C_j$ is circulant with eigenvalues $A_j(\omega)\cos(2^{J-j}\omega)$, whose squares are at most $A_j$. The estimator of Proposition~\ref{prop:lag} takes values in $\{-1,0,1\}$ with second moment $\bar p_j$.
\end{proof}

The Fourier baseline needs a quantum Fourier transform with $O(J^2)$ gates \citep{coppersmith1994,nielsen2010}, no shift register, and no wavelet unitary, and the gains $A_j(\omega_m)$ are computed classically from the observed bin. For the centered wind record of Section~\ref{sec:wind} the ratio of per-shot variances, Fourier to quantum NDWT, ranges from 0.32 to 0.69 for the shares of levels 2 to 8 and from 0.20 to 0.32 for the autocovariances of levels 3 to 8; for the Doppler example it ranges from 0.38 to 0.55 for the shares. For global shift-invariant quadratic statistics the quantum NDWT therefore has no statistical advantage, and the level-share readouts of Proposition~\ref{prop:multinomial} and Proposition~\ref{prop:lag} should be viewed as properties of the construction rather than as recommended estimators.

A second baseline concerns the coherence of the shift register. Every statistic of the data register alone has the same distribution if $\varepsilon$ is drawn uniformly at random, recorded classically, and the single unitary $WS^\varepsilon$ is applied, because $\rho_D$ is exactly that mixture; the joint law (\ref{eq:translate}) is obtained in the same way. Coherence is needed only where branches interfere. One instance is amplitude estimation of $\bar p_j$ on the purification $\ket{\Psi}$ (Section~\ref{sec:ae}); with random shifts the between-branch variance of $p_j^{(\varepsilon)}$ does not decrease with the depth of the Grover circuit, so the $B^{-1}$ rate requires a separate estimation for each of the $2^L$ shifts. The other is the recombination of Section~\ref{sec:shrink}. What the wavelet unitary contributes beyond the Fourier baseline is localization in time and scale, through the diagonal of $\rho_D$ in Theorem~\ref{thm:ti}(a) and the coefficient energies (\ref{eq:translate}), and an inexpensive marking oracle: the reflection $R_j$ below is one multi-controlled $Z$ gate, whereas the Fourier counterpart of $P_j$ is the non-projective observable ${\rm diag}\{A_j(\omega_m)\}$, whose use in amplitude estimation requires reversible evaluation of $A_j$.

\subsection{Hadamard Tests of Reflections}

The Hadamard test estimates $\Re\Tr(V\rho)$ for a unitary $V$ by interfering a controlled application of $V$ with the identity \citep{Aharonov2009,nielsen2010}. Applied to the quantum NDWT, it requires only a controlled reflection, because of the following elementary fact.

\begin{lemma}\label{lem:ctrl}
For unitaries $A$ and $R$, the controlled operator $\ket{0}\bra{0}\otimes I+\ket{1}\bra{1}\otimes A^\dagger RA$ equals $(I\otimes A^\dagger)(\ket{0}\bra{0}\otimes I+\ket{1}\bra{1}\otimes R)(I\otimes A)$.
\end{lemma}

\begin{proof}
On the control value $0$ the right-hand side is $A^\dagger A=I$; on the control value $1$ it is $A^\dagger RA$.
\end{proof}

Take $A=U_{\rm NDWT}$, applied before the test and never controlled, and let $R_j=I-2P_j$ be the reflection about the complement of level $j$. Since $P_j$ fixes the bit $b_j=1$ and the bits above it to zero, $R_j$ is a single multi-controlled $Z$ gate on the $J-j$ most significant qubits, conjugated by $X$ gates on the qubits above $q_j$. The circuit is shown in Fig.~\ref{fig:hadamard}. The ancilla is found in state $\ket{1}$ with probability
\begin{eqnarray}
\Pr(\text{ancilla}=1)=\frac{1-\Tr(R_j\rho_D)}{2}=\bar p_j,
\label{eq:ht}
\end{eqnarray}
so each run is a Bernoulli$(\bar p_j)$ trial. The reflection $R_{i}=I-2\ket{i}\bra{i}$ about a single coefficient gives in the same way the shift-averaged coefficient energy $\bra{i}\rho_D\ket{i}$ of Theorem~\ref{thm:ti}(a); for $L=0$ this is the identity $|w_i|^2=\{1-\bra{y}W^T(I-2\ket{i}\bra{i})W\ket{y}\}/2$ for the decimated coefficient $w_i$.

\begin{figure}[htb]
\centering
\begin{quantikz}[row sep=0.35cm, column sep=0.55cm]
\lstick{$\ket{0}_{\rm anc}$} & \qw & \gate{H} & \ctrl{2} & \gate{H} & \meter{} \\
\lstick{$\ket{0^L}_{\rm shift}$} & \qwbundle{L} & \gate[2]{U_{\rm NDWT}} & \qw & \qw & \qw \rstick{discard} \\
\lstick{$\ket{y}$} & \qwbundle{J} & & \gate{R_j} & \qw & \qw
\end{quantikz}
\caption{\small Hadamard test of the level reflection $R_j=I-2P_j$ on the reduced state. By Lemma~\ref{lem:ctrl} only $R_j$, a multi-controlled $Z$ gate, is controlled; $U_{\rm NDWT}$ includes the preparation $H^{\otimes L}$ of the shift register, which is discarded. The ancilla returns $1$ with probability $\bar p_j$.}
\label{fig:hadamard}
\end{figure}

For a single level, the Hadamard test and the computational-basis measurement have the same per-shot Bernoulli variance, and the latter estimates all levels at once. The value of the reflection lies elsewhere: $R_j$ is exactly the marking oracle that amplitude amplification requires, and the single-qubit readout of (\ref{eq:ht}) is convenient where only one qubit can be measured with high fidelity.

\begin{remark}[A time-domain phase probe does not measure wavelet energy]\label{rem:phase}
The preprint \citet{Vidakovic2025QNDWT} proposed estimating the energy at scale $j$ and position $k$ by a Hadamard test of the diagonal unitary $U_\theta={\rm diag}(e^{i\theta\psi_{j,k}[n]})$, applied to $\ket{y}$ in the time domain. That test returns
$$
\Re\bra{y}U_\theta\ket{y}=\sum_n y_n^2\cos(\theta\psi_{j,k}[n])=1-\frac{\theta^2}{2}\sum_n y_n^2\psi_{j,k}[n]^2+O(\theta^4),
$$
a $\psi^2$-weighted local mass of $y^2$, not the coefficient energy $\langle\psi_{j,k},y\rangle^2$. The two differ in general: if $y$ is the scaling vector supported on the same interval as $\psi_{j,k}$, then $\langle\psi_{j,k},y\rangle=0$ while $\sum_ny_n^2\psi_{j,k}[n]^2>0$. Coefficient energies require the conjugation by $W$ used in (\ref{eq:ht}).
\end{remark}

\subsection{Amplitude Estimation}
\label{sec:ae}

Let $A=U_{\rm NDWT}(H^{\otimes L}\otimes U_y)$ prepare $\ket{\Psi}$ from $\ket{0^{L+J}}$. Since $P_j$ acts on the data register only, $\bar p_j=\|(I\otimes P_j)\ket{\Psi}\|^2$, and the Grover iterate $Q=-A(I-2\ket{0}\bra{0})A^\dagger(I\otimes R_j)$ rotates by the angle $2\theta_j$ with $\sin^2\theta_j=\bar p_j$ in the two-dimensional subspace spanned by the good and bad components of $\ket{\Psi}$. Canonical amplitude estimation with $M$ applications of $Q$ returns $\tilde p_j$ with $|\tilde p_j-\bar p_j|\le 2\pi\sqrt{\bar p_j(1-\bar p_j)}/M+\pi^2/M^2$ with probability at least $8/\pi^2$ \citep{Brassard2002}. For a relative precision $\delta$ in $\bar p_j$, sampling requires about $(1-\bar p_j)/(\bar p_j\delta^2)$ preparations, while amplitude estimation requires on the order of $1/(\delta\sqrt{\bar p_j})$ queries, a quadratic improvement that matters most for small shares at fine or coarse levels \citep{Montanaro2015}. Variants without quantum phase estimation are better suited to near-term devices: maximum-likelihood amplitude estimation combines Bernoulli records at Grover powers $0,1,2,4,\ldots$ \citep{Suzuki2020}, and iterative and low-depth versions trade depth against queries \citep{Grinko2021,GiurgicaTiron2022}. All of them require the inverse $U_y^\dagger$ of the state preparation, so they presuppose coherent access to the data source rather than a supply of copies.

\begin{figure}[htb]
\centering
\includegraphics[width=0.55\linewidth]{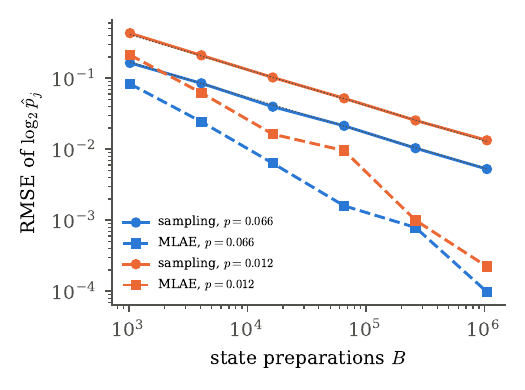}
\caption{\small Root mean squared error of $\log_2\hat p_j$ for computational-basis sampling (circles) and maximum-likelihood amplitude estimation (squares) against the number of state preparations, for shares $p=0.066$ and $p=0.012$; ideal Bernoulli outcomes, 300 replications per point. Dotted lines show the delta-method rate $\{(1-p)/(pB)\}^{1/2}/\ln2$.}
\label{fig:readout}
\end{figure}

Figure~\ref{fig:readout} compares the root mean squared error of $\log_2\hat p_j$ for sampling and for maximum-likelihood amplitude estimation as functions of the number $B$ of state preparations, counting each application of $A$ or $A^\dagger$ as one, for the two shares $\bar p_4=0.066$ and $\bar p_8=0.012$ of the wind data analyzed in Section~\ref{sec:wind}. The simulation draws ideal Bernoulli outcomes with probabilities $\sin^2\{(2m+1)\theta_j\}$; it does not simulate compiled circuits or noise. The schedule uses Grover powers $m\in\{0,1,2,4,\ldots,m_{\max}\}$ with an equal number $N_{\rm sh}\ge100$ of shots per power, the largest schedule whose total query count $N_{\rm sh}\sum(2m+1)$ does not exceed $B$; at $B=2^{20}$ this gives $m_{\max}=2048$ and $N_{\rm sh}=127$, so the longest circuit applies $A$ or $A^\dagger$ $4097$ times. The angle is estimated sequentially: a global grid search over $(0,\pi/2)$ using the first three powers, then for each additional power a local grid search within the smaller of six standard errors and $\pi/(4(2m+1))$ of the current estimate, and finally a global search of the full likelihood over $[\hat\theta/2,3\hat\theta/2]$ at resolution $\pi/\{40(2m_{\max}+1)\}$, which resolves the aliased likelihood peaks that local refinement can miss, followed by a local refinement. An estimate of zero would occur only if all outcomes were zero, and none occurred in these settings. Sampling follows the delta-method rate of Proposition~\ref{prop:multinomial}; maximum-likelihood amplitude estimation follows the rate $B^{-1}$, which for this estimator follows from the Fisher information $4N_{\rm sh}\sum_m(2m+1)^2$ of the schedule \citep{Suzuki2020} rather than from the canonical bound above, and at $B=2^{20}$ it attains errors 54 and 60 times smaller. Whether that gain is useful depends on the statistical problem: Section~\ref{sec:wind} shows that for a single path of length 512 the sampling variability of the path itself dominates the readout error at $n=10^4$ preparations already.

\section{Cycle-Spun Shrinkage by Linear Combination of Unitaries}
\label{sec:shrink}

\subsection{Construction and Exactness}

Cycle spinning estimates a signal $f$ from $x=f+\sigma e$ by averaging a wavelet shrinkage estimator over shifts, $\hat f=2^{-L}\sum_{\varepsilon}S^{-\varepsilon}W^T\delta(WS^\varepsilon x)$ \citep{CoifmanDonoho1995}. For a linear rule $\delta(w)=\Lambda w$ with $\Lambda$ diagonal and $0\le\Lambda\le I$, the estimator is linear in $x$ and can be implemented on the amplitude-encoded state by a linear combination of unitaries \citep{ChildsWiebe2012,Chakraborty2024LCU}. Let $B_\Lambda$ act on the data register and one ancilla $a$ as a block encoding of $\Lambda$, meaning $(\bra{0}_a\otimes I)B_\Lambda(\ket{0}_a\otimes I)=\Lambda$. When $\Lambda=\lambda_sP_s+\sum_j\lambda_jP_j$ is constant within levels, $B_\Lambda$ consists of $L+1$ rotations $R_y(2\arccos\lambda_j)$ on the ancilla, each controlled by the leading-one pattern that defines $P_j$; a general diagonal $\Lambda$ requires one rotation per coordinate, which a uniformly controlled rotation realizes with $O(N)$ gates \citep{Mottonen2004}.

\begin{figure}[htb]
\centering
\begin{quantikz}[row sep=0.35cm, column sep=0.45cm]
\lstick{$\ket{0^L}_{\rm shift}$} & \qwbundle{L} & \gate{H^{\otimes L}} & \ctrl{2} & \qw & \qw & \qw & \ctrl{2} & \gate{H^{\otimes L}} & \meter{} \rstick{$0^L$} \\
\lstick{$\ket{0}_{a}$} & \qw & \qw & \qw & \qw & \gate[2]{B_\Lambda} & \qw & \qw & \qw & \meter{} \rstick{$0$} \\
\lstick{$\ket{y}$} & \qwbundle{J} & \qw & \gate{S^{\varepsilon}} & \gate{W} & & \gate{W^{T}} & \gate{S^{-\varepsilon}} & \qw & \qw \rstick{$\ket{\hat f}/\|\hat f\|$}
\end{quantikz}
\caption{\small Cycle-spun linear shrinkage as a linear combination of unitaries. The block encoding $B_\Lambda$ applies the diagonal shrinkage operator in the wavelet domain on every shift branch; the inverse transform, inverse controlled shift, and a second $H^{\otimes L}$ recombine the branches. Postselection of the shift register on $0^L$ and of the ancilla on $0$ heralds the cycle-spinning estimate.}
\label{fig:lcu}
\end{figure}

\begin{theorem}\label{thm:lcu}
In the circuit of Fig.~\ref{fig:lcu}, the component of the output with the shift register in $\ket{0^L}$ and the ancilla in $\ket{0}$ is $\ket{0^L}\ket{0}\otimes T_\Lambda\ket{y}$ with
\begin{eqnarray}
T_\Lambda=\frac{1}{2^L}\sum_{\varepsilon=0}^{2^L-1}S^{-\varepsilon}W^T\Lambda WS^{\varepsilon}.
\label{eq:tlambda}
\end{eqnarray}
Hence the heralded data state is $\hat f/\|\hat f\|$, where $\hat f=T_\Lambda x$ is the cycle-spinning linear shrinkage estimator, and the heralding probability is $\pi_\Lambda=\|\hat f\|^2/\|x\|^2$.
\end{theorem}

\begin{proof}
After the first Hadamard layer, the controlled shift, and $W$, the state is
$$
2^{-L/2}\sum_\varepsilon\ket{\varepsilon}\ket{0}_aWS^\varepsilon\ket{y}.
$$
The block encoding maps $\ket{0}_a\otimes v$ to $\ket{0}_a\otimes\Lambda v+\ket{1}_a\otimes v'$ for some $v'$. The inverse transform and inverse controlled shift act within each branch, giving
$$
2^{-L/2}\sum_\varepsilon\ket{\varepsilon}\big\{\ket{0}_aS^{-\varepsilon}W^T\Lambda WS^\varepsilon\ket{y}+\ket{1}_a(\cdot)\big\}.
$$ Since $\bra{0^L}H^{\otimes L}\ket{\varepsilon}=2^{-L/2}$, projecting the shift register onto $\ket{0^L}$ after the second Hadamard layer and the ancilla onto $\ket{0}$ leaves $T_\Lambda\ket{y}$, whose squared norm is the heralding probability.
\end{proof}

Theorem~\ref{thm:lcu} is an exact implementation of the finite shift average (\ref{eq:tlambda}) for every diagonal $\Lambda$. Whether that average is translation equivariant depends on $\Lambda$ and is settled below. The squared norm of the estimate, which the amplitude encoding cannot carry, is recovered from the heralding probability as $\|\hat f\|^2=\|x\|^2\pi_\Lambda$, and $\pi_\Lambda$ is estimated by the frequency of successful heralds. If $\pi_\Lambda$ is small, amplitude amplification raises it with $O(\pi_\Lambda^{-1/2})$ rounds. The gate-level statevector of the circuit agreed with the classical cycle-spinning estimator to within $4\times10^{-16}$ for Haar and db2 transforms with level-wise and general diagonal weights.

For level-wise weights the estimator has a transparent spectral form. Let $T_j=2^{-L}\sum_\varepsilon S^{-\varepsilon}W^TP_jWS^\varepsilon$ and define $T_s$ likewise.

\begin{corollary}\label{cor:filter}
Each $T_j$ equals $2^{j-J}\sum_{\tau\in\mathbb{Z}_N}g_{j,0}(\cdot-\tau)\,g_{j,0}(\cdot-\tau)^T$, a symmetric circulant matrix with eigenvalues $A_j(\omega)=2^{j-J}|\hat g_{j,0}(\omega)|^2\ge0$ at the Fourier frequencies $\omega=2\pi m/N$, and similarly for $T_s$. The gains form a partition of unity, $A_s(\omega)+\sum_jA_j(\omega)=1$. Consequently, for $\Lambda=\lambda_sP_s+\sum_j\lambda_jP_j$, the operator $T_\Lambda$ is the zero-phase linear time-invariant filter with transfer function
\begin{eqnarray}
H(\omega)=\lambda_sA_s(\omega)+\sum_{j=J-L}^{J-1}\lambda_jA_j(\omega),
\label{eq:transfer}
\end{eqnarray}
a convex combination of the wavelet band gains, and $\pi_\Lambda=\sum_m H(\omega_m)^2|\hat y(\omega_m)|^2/N$. In particular $T_\Lambda$ commutes with every circular shift, and so does the estimator when the level weights are computed from the shift-invariant shares.
\end{corollary}

\begin{proof}
By (\ref{eq:branch}) and Lemma~\ref{lem:coverage}, $\sum_\varepsilon S^{-\varepsilon}W^TP_jWS^\varepsilon=\sum_{\varepsilon,k}(S^{-\varepsilon}g_{j,k})(S^{-\varepsilon}g_{j,k})^T$ sums the outer products of all $N$ translates of $g_{j,0}$, each $2^{L-J+j}$ times. A sum over all translates of an outer product is circulant with eigenvalues $|\hat g_{j,0}|^2$. The partition of unity follows from $\sum_jT_j+T_s=2^{-L}\sum_\varepsilon S^{-\varepsilon}W^TWS^\varepsilon=I$, and the heralding probability follows from Parseval's identity.
\end{proof}

Corollary~\ref{cor:filter} explains the advantage of the translation-invariant estimator over its decimated counterpart for linear rules: the decimated operator $W^T\Lambda W$ is periodically time varying and aliases between neighboring bands, whereas $T_\Lambda$ is time invariant with the smooth transfer function (\ref{eq:transfer}). It also shows what level-wise weights cannot do: a filter cannot adapt to local features. Coordinatewise weights can, but equivariance is then no longer automatic.

\begin{proposition}\label{prop:equiv}
(a) If $\Lambda$ is not constant within levels, $T_\Lambda$ need not commute with $S$. If the coordinatewise weights are computed from the diagonal of $\rho_D$, the resulting estimator satisfies $\hat f(S^mx)=S^m\hat f(x)$ for every multiple $m$ of $2^L$, but not in general for other $m$. (b) Replace $B_\Lambda$ in Fig.~\ref{fig:lcu} by $\sum_\varepsilon\ket{\varepsilon}\bra{\varepsilon}\otimes B_{\Lambda_\varepsilon}$, controlled by the shift register, with $(\Lambda_\varepsilon)_{2^j+k}=\lambda_j[2^{J-j}k-\varepsilon]$ for weight sequences $\lambda_j$ on $\mathbb Z_N$ with $0\le\lambda_j\le1$, and scaling weights one. The heralded state is then proportional to
\begin{eqnarray}
\hat f=\sum_{j=J-L}^{J-1}2^{j-J}\sum_{\tau\in\mathbb Z_N}\lambda_j[\tau]\,d_j[\tau]\,g_{j,0}(\cdot-\tau)+2^{-L}\sum_{\tau\in\mathbb Z_N}a[\tau]\,\phi_{J-L,0}(\cdot-\tau),
\label{eq:tauest}
\end{eqnarray}
with heralding probability $\|\hat f\|^2/\|x\|^2$. If the weights are computed from $x$ by an equivariant rule, $\lambda_j(S^mx)[\tau]=\lambda_j(x)[\tau-m]$, then $\hat f(S^mx)=S^m\hat f(x)$ for every $m\in\mathbb Z_N$.
\end{proposition}

\begin{proof}
(a) Shifting the input by $2^L$ permutes the coefficients cyclically within each level, $WS^{2^L}=\Pi W$, so the diagonal of $\rho_D$ and the weights computed from it are permuted by $\Pi$, and $T_{\Pi\Lambda\Pi^T}=S^{2^L}T_\Lambda S^{-2^L}$, which gives equivariance for multiples of $2^L$. Remark~\ref{rem:counter} gives a counterexample for $m=1$. (b) The proof of Theorem~\ref{thm:lcu} applies branch by branch and gives $2^{-L}\sum_\varepsilon S^{-\varepsilon}W^T\Lambda_\varepsilon WS^\varepsilon x$. By (\ref{eq:branch}), branch $\varepsilon$ contributes $2^{-L}\lambda_j[\tau]d_j[\tau]\,g_{j,0}(\cdot-\tau)$ at index $2^j+k$ with $\tau=2^{J-j}k-\varepsilon$, and Lemma~\ref{lem:coverage} counts each $\tau$ exactly $2^{L-J+j}$ times at level $j$ and once in the scaling band, which gives (\ref{eq:tauest}). Shifting $x$ by $m$ shifts every $d_j$ and $a$ by $m$, and, by assumption, every $\lambda_j$; each sum in (\ref{eq:tauest}) then shifts by $m$.
\end{proof}

\begin{remark}[A four-point counterexample]\label{rem:counter}
Take the Haar transform with $N=4$, $L=1$, scaling weights one, and the index-wise rule $\lambda_i=(1-c\sigma^2/\bra{i}\rho_D\ket{i})_+$ with $c\sigma^2=0.3$ and $\|x\|=1$. For $x=e_0$ the two detail energies are $(0.5,0)$ and the weights $(0.4,0)$, giving $\hat f(e_0)=(0.70,0.15,0,0.15)$. For $Sx=e_1$ the energies are $(0.25,0.25)$ and both weights vanish, giving $\hat f(e_1)=(0.25,0.50,0.25,0)\ne S\hat f(e_0)=(0.15,0.70,0.15,0)$. The failure is structural and is not repaired by more shots: Lemma~\ref{lem:coverage} counts occurrences of each translate, but with index-wise weights different occurrences of the same translate receive different weights. We thank a referee for this example.
\end{remark}

\begin{remark}[Channels and heralding]\label{rem:channel}
The preprint \citet{Vidakovic2025QNDWT} described shrinkage as a completely positive trace-preserving map in the wavelet domain. A trace-preserving map cannot shrink an amplitude-encoded signal, since the trace is the normalized energy. With the ancilla traced out, the block encoding induces $\rho\mapsto K_0\rho K_0^\dagger+K_1\rho K_1^\dagger$ with the diagonal Kraus operators $K_0=\Lambda$ and $K_1=(I-\Lambda^2)^{1/2}$. Because $\lambda_i^2+(1-\lambda_i^2)=1$, this channel preserves every wavelet-basis population and multiplies the coherence between coordinates $i$ and $l$ by $\lambda_i\lambda_l+\{(1-\lambda_i^2)(1-\lambda_l^2)\}^{1/2}\le1$; it is a dephasing-type channel, and complete dephasing would produce ${\rm diag}(w_i^2)$, the energy distribution of Theorem~\ref{thm:ti}(a), not an estimate of $f$. Shrinkage is the trace-decreasing operation $\rho\mapsto K_0\rho K_0^\dagger$ selected by the ancilla outcome $0$ \citep{stinespring1955,kraus1971,nielsen2010}, and its probability is the informative quantity $\pi_\Lambda$ of Theorem~\ref{thm:lcu}.
\end{remark}

\subsection{Data-Adaptive Weights by Measurement and Feed-Forward}

A fixed quantum circuit acts linearly on the amplitudes of a single copy, so coefficientwise nonlinear rules such as hard or soft thresholding cannot be applied by any fixed operation on $\ket{y}$. Nonlinear transformations of amplitudes are possible with repeated coherent queries to $U_y$ through quantum singular value transformation \citep{Gilyen2019,GuoMitaraiFujii2024,RattewRebentrost2023}, but their cost depends on the dimension and on the normalization of the amplitudes, and their use for wavelet thresholding is an open question. A cheaper route to adaptivity, in the spirit of empirical Bayes and of classical two-stage smoothers, is to estimate the weights from measurements of $\rho_D$ and feed them forward into $B_\Lambda$. The overall estimator is then nonlinear in $x$ through the weights while each heralded run remains linear.

Suppose $x=f+\sigma e$ with $e\sim N(0,I_N)$ and $\sigma$ known or estimated separately. Since each NDWT filter has unit norm, $E\,d_j[\tau]^2=d_j^f[\tau]^2+\sigma^2$, and Theorem~\ref{thm:ti} gives
$$
E\big(\|x\|^2\bar p_j\big)=\bar\theta_j^2+2^j\sigma^2,\qquad \bar\theta_j^2=2^{j-J}\sum_\tau d_j^f[\tau]^2 .
$$
The branchwise oracle weights $\lambda_j^{\rm bw}=\bar\theta_j^2/(\bar\theta_j^2+2^j\sigma^2)$ minimize the average risk of the shrunken coefficients of level $j$ before reconstruction and averaging. They do not in general minimize the risk of the averaged output, because the band gains overlap. By Corollary~\ref{cor:filter} that risk is
\begin{eqnarray}
E\|T_\lambda x-f\|^2=\frac1N\sum_m\{H(\omega_m)-1\}^2|\hat f(\omega_m)|^2+\sigma^2\sum_mH(\omega_m)^2,
\label{eq:risk}
\end{eqnarray}
with $\hat f$ the unnormalized discrete Fourier transform and $H$ given by (\ref{eq:transfer}) with $\lambda_s=1$, a convex quadratic in $(\lambda_{J-L},\ldots,\lambda_{J-1})$. Its minimizer over $[0,1]^L$ is the risk oracle $\lambda^{\rm risk}$, computed by box-constrained least squares. Replacing the unknown energies by unbiased estimates gives the level-wise rule
\begin{eqnarray}
\hat\lambda_j=\Big(1-\frac{2^j\sigma^2}{\|x\|^2\hat p_j}\Big)_+,\qquad J-L\le j\le J-1,
\label{eq:js}
\end{eqnarray}
a James--Stein-type positive-part plug-in rule. It differs from the classical positive-part James--Stein rule for a single Gaussian block of dimension $2^j$, which subtracts $(2^j-2)\sigma^2$ from the block energy; here the energy is shift averaged and the difference matters only for the coarsest levels. The rule uses the multinomial estimates of Proposition~\ref{prop:multinomial} and $L$ controlled rotations.

For local adaptivity, Proposition~\ref{prop:equiv} dictates that the weights be indexed by the NDWT translate. The joint record (\ref{eq:translate}) of both registers gives unbiased estimates $\hat e_j[\tau]=2^{J-j}\|x\|^2n_{j,\tau}/n$ of $d_j[\tau]^2$, and the local empirical Wiener rule is
\begin{eqnarray}
\hat\lambda_j[\tau]=\Big(1-\frac{c\,\sigma^2}{\bar e_j[\tau]}\Big)_+,\qquad \bar e_j[\tau]=\frac{1}{w}\sum_{u=-w/2}^{w/2-1}\hat e_j[\tau+u],
\label{eq:ew}
\end{eqnarray}
a moving average over a window of $w=2^L$ translates with fixed offsets. With exact energies the rule is equivariant, so by Proposition~\ref{prop:equiv}(b) the estimator (\ref{eq:tauest}) satisfies $\hat f(S^mx)=S^m\hat f(x)$ for every $m$; with estimated energies the counts of the shifted input have the shifted distribution, so equivariance holds in distribution. The rule is a translation-equivariant relative of block thresholding \citep{cai1999}, of empirical Wiener filtering in the wavelet domain \citep{Ghael1997}, and of the nonnegative garrote \citep{Gao1998}. The constant $c=1$ corresponds to unbiased estimation of the local signal energy; results for $c=2$ are exploratory, because that value was selected after inspecting results on the same test functions. For comparison we also report the index-wise rule that applies $(1-\sigma^2/\{\|x\|^2\bra{i}\rho_D\ket{i}\})_+$ to coordinate $i$ in every branch. By Proposition~\ref{prop:equiv}(a) it is equivariant only for shifts by multiples of $2^L$, so it is a finite-shift estimator, not a translation-invariant one.

The block encoding of Proposition~\ref{prop:equiv}(b) is controlled by the $J+L$ qubits of both registers, but its angles take at most $LN$ distinct values. A reversible subtraction computes $\tau=2^{J-j}k-\varepsilon$ into a work register with $O(J)$ gates, and a table lookup of $LN$ angles followed by $b$ bit-controlled rotations costs $O(LN)$ gates for $b$-bit angles \citep{Babbush2018}, compared with $O(N)$ for the index-wise rule and $O(L^2)$ for the level-wise rule. Coordinatewise adaptivity therefore costs gates linear in $N$, the same order as preparing a general input state.

\subsection{Simulation Study}

We used the four test functions of \citet{donoho1994}, Doppler, HeaviSine, Bumps, and Blocks, sampled at $N=1024$ ($J=10$), scaled to unit standard deviation, and observed in Gaussian noise with signal-to-noise ratios $7$ and $3$, that is, $\sigma=1/7$ and $\sigma=1/3$. All estimators used the Daubechies transform with four vanishing moments and depth $L=5$, so that the levels are $W_5,\ldots,W_9$ and $V_5$. The error criterion is $\|\hat f-f\|^2/N$ for the classical vector $\hat f$ that the heralded state encodes, with its norm taken from the heralding probability, computed exactly in the simulation and reported divided by $\sigma^2$. It is not the error of a vector read out from a device; Section~\ref{sec:cost} gives the cost of such a readout. Table~\ref{tab:shrinkA} compares level-wise estimators: the decimated estimator with branchwise oracle weights computed from the energies of the same shift, the cycle-spun estimator (\ref{eq:tlambda}) with branchwise oracle weights and with risk-oracle weights, and the rule (\ref{eq:js}) with shares estimated from $n=10^4$ multinomial preparations and with exact shares. Table~\ref{tab:shrinkB} compares local rules: the index-wise finite-shift rule with exact energies, the equivariant rule (\ref{eq:ew}) with $c=1$ from $n=10^5$ and $n=10^6$ joint measurements and with exact energies, the exploratory $c=2$ with exact energies, and, as a classical benchmark outside the scope of the linear construction, cycle-spinning hard thresholding at the universal threshold $\sigma\sqrt{2\log N}$ \citep{donoho1994,CoifmanDonoho1995}. Each entry averages 100 replications; the largest Monte Carlo standard error of an entry is 0.0036.

\begin{table}[htb]
\centering
{\footnotesize
\begin{tabular}{lc|ccc|cc|c}
\toprule
 & & \multicolumn{3}{c|}{Oracle weights} & \multicolumn{2}{c|}{Rule (\ref{eq:js})} & Herald\\
Function & SNR & Dec., bw & TI, bw & TI, risk & $n=10^4$ & exact & prob.\\
\midrule
Doppler & 7 & 0.384 & 0.353 & 0.350 & 0.360 & 0.355 & 0.97 \\
 & 3 & 0.219 & 0.205 & 0.205 & 0.209 & 0.208 & 0.89 \\
HeaviSine & 7 & 0.107 & 0.103 & 0.102 & 0.112 & 0.106 & 0.98 \\
 & 3 & 0.053 & 0.049 & 0.049 & 0.052 & 0.051 & 0.91 \\
Bumps & 7 & 0.767 & 0.732 & 0.719 & 0.733 & 0.733 & 0.97 \\
 & 3 & 0.490 & 0.458 & 0.453 & 0.462 & 0.461 & 0.87 \\
Blocks & 7 & 0.605 & 0.589 & 0.587 & 0.593 & 0.590 & 0.98 \\
 & 3 & 0.313 & 0.300 & 0.299 & 0.304 & 0.302 & 0.92 \\
\bottomrule
\end{tabular}
}
\caption{\small Level-wise shrinkage: mean squared error divided by $\sigma^2$, db4, $N=1024$, $L=5$, 100 replications. bw, branchwise oracle weights; risk, minimizer of (\ref{eq:risk}); Dec., decimated; TI, cycle-spun over $2^L$ shifts, translation equivariant by Corollary~\ref{cor:filter}. The last column is the average heralding probability of the exact-share rule.}
\label{tab:shrinkA}
\end{table}

\begin{table}[htb]
\centering
{\footnotesize\setlength{\tabcolsep}{4pt}
\begin{tabular}{lc|c|cccc|c|c}
\toprule
 & & Index-wise & \multicolumn{4}{c|}{Equivariant local rule (\ref{eq:ew})} & TI hard & Herald\\
Function & SNR & finite shift & $n=10^5$ & $n=10^6$ & exact & $c=2$ & (classical) & prob.\\
\midrule
Doppler & 7 & 0.134 & 0.144 & 0.141 & 0.140 & 0.091 & 0.119 & 0.98 \\
 & 3 & 0.116 & 0.123 & 0.123 & 0.122 & 0.079 & 0.107 & 0.90 \\
HeaviSine & 7 & 0.098 & 0.111 & 0.107 & 0.106 & 0.064 & 0.066 & 0.98 \\
 & 3 & 0.083 & 0.092 & 0.091 & 0.091 & 0.046 & 0.049 & 0.91 \\
Bumps & 7 & 0.331 & 0.359 & 0.356 & 0.356 & 0.341 & 0.212 & 0.98 \\
 & 3 & 0.260 & 0.278 & 0.277 & 0.277 & 0.270 & 0.198 & 0.90 \\
Blocks & 7 & 0.305 & 0.336 & 0.331 & 0.331 & 0.359 & 0.201 & 0.99 \\
 & 3 & 0.223 & 0.239 & 0.237 & 0.237 & 0.239 & 0.212 & 0.93 \\
\bottomrule
\end{tabular}
}
\caption{\small Local shrinkage: mean squared error divided by $\sigma^2$, same design as Table~\ref{tab:shrinkA}. The index-wise rule is equivariant only under shifts by multiples of $2^L$ (Proposition~\ref{prop:equiv}(a)); the equivariant rule uses translate-indexed weights with $c=1$ unless stated; $c=2$ is exploratory. The last column is the average heralding probability of the exact equivariant rule.}
\label{tab:shrinkB}
\end{table}

\begin{figure}[htb]
\centering
\includegraphics[width=\linewidth]{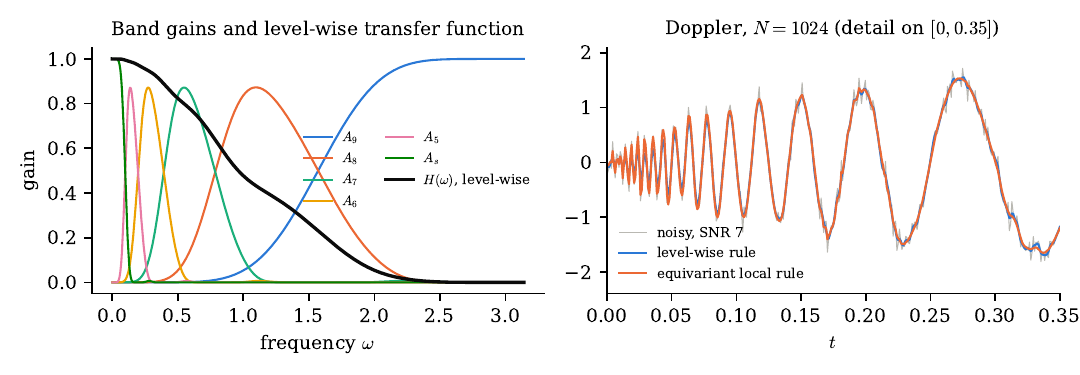}
\caption{\small Left: band gains $A_j(\omega)$, $j=5,\ldots,9$, and $A_s(\omega)$ of the db4 transform with $N=1024$ and $L=5$, which sum to one, and the transfer function $H(\omega)$ of the level-wise rule (\ref{eq:js}) for a noisy Doppler realization at signal-to-noise ratio 7. Right: the noisy data and the level-wise rule (\ref{eq:js}) and equivariant local rule (\ref{eq:ew}), $c=1$, estimates on $[0,0.35]$.}
\label{fig:shrink}
\end{figure}

Five conclusions emerge. First, the cycle-spun level-wise estimator improves on the decimated one in every case, by 2.6 to 8.1 percent, consistent with Corollary~\ref{cor:filter}. Second, the branchwise oracle is nearly risk optimal: its risk exceeds that of the risk oracle by at most 0.013, and the rule (\ref{eq:js}) with exact shares is within 0.014 of the risk oracle in every case. Third, measurement noise in the weights is small but not zero. Paired over replications, the excess risk of the level-wise rule with $10^4$ preparations over exact shares lies between $-0.001$ and $0.006$, with paired standard errors at most $0.001$. For the equivariant local rule it lies between $0.001$ and $0.006$ with $10^5$ joint measurements and below $0.001$ with $10^6$, with paired standard errors at most $0.0006$. These differences are detectable, but they are an order of magnitude smaller than the differences between methods. Fourth, local adaptivity matters. At signal-to-noise ratio 7 the equivariant local rule reduces the risk of the level-wise rule by 60, 52, and 44 percent for Doppler, Bumps, and Blocks; its weakest case is HeaviSine at signal-to-noise ratio 3, where the smooth signal favors pooling over levels. The finite-shift index-wise rule has risk lower by 0.006 to 0.026. The two rules differ both in equivariance and in their effective windows, so this experiment does not attribute the gap to either difference. Fifth, nonlinear thresholding remains superior for signals with jumps and narrow peaks, with risks of about 0.20 to 0.21 against 0.24 to 0.36 for the equivariant local rule with $c=1$ on Bumps and Blocks; the exploratory $c=2$ matches hard thresholding for Doppler and HeaviSine but not for Bumps and Blocks. The heralding probabilities exceed 0.87 throughout, so amplitude amplification is unnecessary in this regime. Figure~\ref{fig:shrink} shows the band gains $A_j(\omega)$, the transfer function (\ref{eq:transfer}) of the level-wise rule for one noisy Doppler realization, and the level-wise and local estimates.

\subsection{Cost of Learning Weights and of Producing the Output}
\label{sec:cost}

The two stages have different costs and should not be conflated. Both use the classical inputs $\|x\|^2$, retained when $\ket{y}$ is prepared, and $\sigma^2$, assumed known or estimated separately. Learning the weights consumes fresh copies: each is one preparation of $\ket{y}$ followed by $U_{\rm NDWT}$ on $J+L$ qubits, plus one work qubit for the controlled increments, and a measurement of the data register for the level-wise rule or of both registers for the local rule. The simulations used $n=10^4$ copies for the level-wise rule and $10^5$ to $10^6$ for the local rule, whose $LN=5120$ coefficient energies are averaged over windows of $2^L=32$ translates. Producing the output costs one run of the circuit of Fig.~\ref{fig:lcu} per attempt, with $J+L+1$ qubits plus the work registers of the block encoding, namely a $J$-qubit translate register and the lookup ancillas for the local rule, and succeeds with probability $\pi_\Lambda$; the expected number of attempts per heralded state is $1/\pi_\Lambda$, below $1.2$ in Tables~\ref{tab:shrinkA} and \ref{tab:shrinkB}. The product of this stage is a quantum state. Returning the classical vector $\hat f$ requires state tomography of the heralded state, which for error $\eta$ in the normalized vector needs on the order of $N/\eta^2$ heralded copies \citep{Haah2017}, and the norm $\|\hat f\|=\|x\|\pi_\Lambda^{1/2}$ from the heralding frequency. Tomography determines the normalized state only up to a global phase, so for a real signal the overall sign of $\hat f$ needs a convention or a phase reference. A natural convention is available: every estimator of this section has the form $\hat f=Kx$ with $K\succeq0$, by (\ref{eq:tlambda}) with $\Lambda\succeq0$ or, for (\ref{eq:tauest}), by Proposition~\ref{prop:direct} below, so $\langle x,\hat f\rangle\ge0$. Applying the convention requires the sign of the overlap between the input state and the tomographic estimate, which one Hadamard test using the preparation unitary $U_y$ and a preparation of the estimate provides. A classical denoised vector therefore costs order $N$ copies and forfeits any advantage; the construction is useful when the heralded state is consumed by further quantum processing or by estimation of a few functionals of $\hat f$.

\subsection{Equivalent Output Maps and What the Shift Register Contributes}

The heralded estimators of this section can be written without reference to the shift register. Let $C:\mathbb R^N\to\mathbb R^{(L+1)N}$ be the energy-normalized NDWT analysis map,
$$
Cx=\big(2^{(j-J)/2}d_j\big)_{j=J-L}^{J-1}\oplus2^{-L/2}a .
$$

\begin{proposition}\label{prop:direct}
(a) $C$ is an isometry, and the estimator (\ref{eq:tauest}) equals $C^\top DCx$ with $D=\operatorname{diag}(\lambda_{J-L},\ldots,\lambda_{J-1},\mathbf 1)$. Consequently any unitary embedding of $C$, followed by a block encoding of $D$ and the inverse embedding, heralds the same state with the same probability $\|C^\top DCx\|^2/\|x\|^2$, and a computational-basis measurement of the embedded state $C\ket{y}$ has the law (\ref{eq:translate}). (b) For level-wise weights, $T_\lambda=F^\dagger\operatorname{diag}\{H(\omega_m)\}F$ by Corollary~\ref{cor:filter}, so a block encoding of $\operatorname{diag}\{H(\omega_m)\}$ between a quantum Fourier transform and its inverse heralds the same state with the same probability. (c) In the circuit of Fig.~\ref{fig:lcu}, replacing the first $H^{\otimes L}$ by a unitary $V_\mu$ with $V_\mu\ket{0^L}=\sum_\varepsilon\mu_\varepsilon^{1/2}\ket\varepsilon$ and the second by $V_\mu^\dagger$ heralds $\sum_\varepsilon\mu_\varepsilon S^{-\varepsilon}W^T\Lambda_\varepsilon WS^\varepsilon x$ for any probability vector $\mu$.
\end{proposition}

\begin{proof}
(a) $\|Cx\|^2=\|x\|^2$ is the statement $\bar p_s+\sum_j\bar p_j=1$ of Theorem~\ref{thm:ti}, and $C^\top DCx$ is (\ref{eq:tauest}) term by term, since $C^\top$ synthesizes $\sum_\tau2^{(j-J)/2}v_j[\tau]g_{j,0}(\cdot-\tau)$ at level $j$. The law of $C\ket{y}$ is $2^{j-J}d_j[\tau]^2/\|x\|^2$, which is (\ref{eq:translate}). (b) is Corollary~\ref{cor:filter}. (c) follows from the proof of Theorem~\ref{thm:lcu}, with $\bra{0^L}V_\mu^\dagger\ket\varepsilon\bra\varepsilon V_\mu\ket{0^L}=\mu_\varepsilon$ in place of $2^{-L}$.
\end{proof}

Parts (a) and (b) show that the output maps of Theorem~\ref{thm:lcu} and Proposition~\ref{prop:equiv} are not specific to the shift-register architecture. A unitary embedding of the nondecimated filter bank such as that of \citet{nakahira2021}, with the per-level normalization of $C$, implements the same contraction, and it indexes the weights by $(j,\tau)$ directly, without the subtraction that computes $\tau$ from $(\varepsilon,k)$. Its block encoding of $D$ has $LN$ distinct angles, the same order as ours. For level-wise weights the Fourier route needs only a quantum Fourier transform, but its block encoding must evaluate $H(\omega)=\sum_j\lambda_jA_j(\omega)$ reversibly from products of trigonometric polynomials of the filter, whereas in the wavelet domain the same operator is $L+1$ rotations controlled by leading-one patterns. The heralding probabilities are identical in all three routes, so the comparison among them is one of gate cost, which we have not compiled.

What the shift-register construction contributes is therefore not a different output map but access to the individual decimated branches within one circuit. Measuring the shift register exposes the decimated analysis of every branch and supports the test of shift dependence of Section~\ref{sec:measurement}. Branch-dependent operations, such as the weights $\Lambda_\varepsilon$, act on the branches as such. Part (c) shows that the recombination itself is adjustable: a nonuniform shift state $V_\mu$ yields a weighted cycle spinning that interpolates between a single decimated estimator, with $\mu$ a point mass, and the uniform average of Theorem~\ref{thm:lcu}. In the analysis-synthesis route these branch-level operations would have to be reconstructed from the nondecimated coefficients.

\section{Scaling Analysis of Turbulent Wind Velocity}
\label{sec:wind}

\begin{figure}[htb]
\centering
\includegraphics[width=\linewidth]{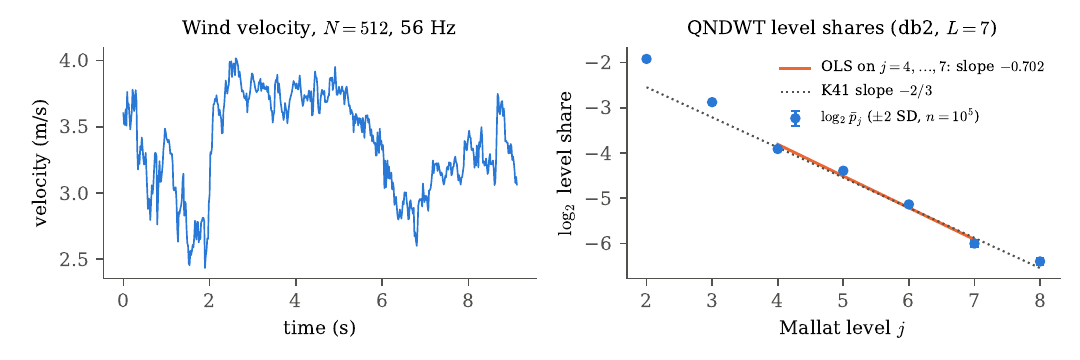}
\caption{\small Left: wind velocity, $N=512$, sampled at 56~Hz. Right: $\log_2$ level shares of the db2 quantum NDWT with $L=7$, with $\pm2$ standard deviation bars for $n=10^5$ preparations from Proposition~\ref{prop:multinomial}, the least-squares line on levels $4$ to $7$, and a line of slope $-2/3$.}
\label{fig:wind}
\end{figure}

This section illustrates the level-share spectrum and the within-level autocorrelations on one record; by Proposition~\ref{prop:qft} the same global statistics are also available from a Fourier-basis measurement. We use a segment of $N=512$ longitudinal wind velocities sampled at 56~Hz in the atmospheric surface layer \citep{VidakovicKatulAlbertson2000} (Fig.~\ref{fig:wind}, left), the segment distributed with the accompanying notebooks, whose first values are $3.602$, $3.520$, and $3.515$~m/s and whose mean and standard deviation are $3.38$ and $0.37$~m/s. In the inertial subrange, Kolmogorov's theory predicts a velocity spectrum proportional to $\omega^{-5/3}$, the spectrum of fractional Brownian motion with Hurst exponent $H=1/3$ \citep{Kolmogorov1941}. For fractional Brownian motion and a wavelet with at least one vanishing moment, the expected squared NDWT coefficient at level $j$ is proportional to $2^{-j(2H+1)}$ in Mallat's indexing \citep{Flandrin1992,AbryVeitch1998}, and since (\ref{eq:shares}) weights the $N$ squared coefficients of level $j$ by $2^{j-J}$,
$$
\log_2\bar p_j=c-2Hj
$$
up to fluctuations, so the log-share spectrum has slope $-2H$, or $-2/3$ under the theory.

We removed the sample mean of 3.38~m/s classically before loading, for the reason given after Proposition~\ref{prop:multinomial}, and computed the quantum NDWT with db2 and $L=7$, so that the levels are $W_2,\ldots,W_8$. The right panel of Fig.~\ref{fig:wind} displays $\log_2\bar p_j$. The fitting range $j=4,\ldots,J-2$, that is, levels 4 to 7, was fixed in advance by the rule used in the companion analysis of Hurst exponent estimation: it omits the two coarsest levels, which hold four and eight decimated coefficients per shift and are expected to extend beyond the inertial subrange, and the finest level. Ordinary least squares on these levels gives slope $-0.702$ and $\hat H=0.351$. This is an illustration from one record and four fitted levels rather than an estimate with stated precision. Table~\ref{tab:sens} shows its sensitivity to the fitting range and to the boundary treatment: over adjacent ranges and for both the periodic transform and the mirror extension, which costs one additional data qubit, $\hat H$ lies between 0.33 and 0.40. The coarsest levels lie above the fitted line and the finest level lies below it.

Two sources of variability affect $\hat H$, and the comparison between them determines the measurement budget. The readout error from $n$ multinomial preparations, obtained by simulating the counts, gives standard deviations of $\hat H$ of $0.020$, $0.0064$, and $0.0020$ for $n=10^4$, $10^5$, and $10^6$. The sampling variability of the path itself, obtained from 1000 simulated paths of fractional Brownian motion with $H=1/3$ and $N=512$ analyzed identically with exact shares, gives a standard deviation of $0.072$ and a mean of $0.299$ (Table~\ref{tab:sens}); the downward bias is examined below. Readout noise is therefore negligible relative to sampling variability once $n\approx10^4$, and the quadratic speedup of amplitude estimation brings no statistical benefit for this problem. A full treatment of Hurst exponent estimation from quantum NDWT shares, with mirror extension, calibrated generalized least squares, and asymptotic theory, is the subject of a companion paper.

\begin{table}[htb]
\centering
{\footnotesize
\begin{tabular}{c|cc|cc|cc}
\toprule
 & \multicolumn{2}{c|}{Wind record, $\hat H$} & \multicolumn{2}{c|}{fBm, periodic} & \multicolumn{2}{c}{fBm, mirror}\\
Levels & periodic & mirror & bias & SD & bias & SD\\
\midrule
3--7 & 0.374 & 0.392 & $-0.025$ & 0.074 & $-0.053$ & 0.070 \\
4--7 & 0.351 & 0.359 & $-0.034$ & 0.072 & $-0.061$ & 0.068 \\
5--7 & 0.403 & 0.394 & $-0.054$ & 0.074 & $-0.080$ & 0.072 \\
3--8 & 0.352 & 0.367 & $-0.069$ & 0.058 & $-0.095$ & 0.052 \\
4--8 & 0.329 & 0.337 & $-0.091$ & 0.054 & $-0.117$ & 0.049 \\
2--8 & 0.373 & 0.387 & $-0.054$ & 0.063 & $-0.084$ & 0.058 \\
\bottomrule
\end{tabular}
}
\caption{\small Sensitivity of the least-squares estimate of $H$ to the fitting range (Mallat levels of the original record) and to boundary treatment, db2. Wind: estimates from the centered record. fBm: bias and standard deviation of the same estimator over 1000 simulated centered paths with $H=1/3$ and $N=512$, exact shares.}
\label{tab:sens}
\end{table}

The simulation in Table~\ref{tab:sens} also bears on the downward bias. It is not caused by the periodic boundary: the mirror extension, which removes the boundary jump, increases it, from $-0.034$ to $-0.061$ for levels 4 to 7. Fitting the logarithms of the exact expected energies of a centered fractional Brownian path, computed from its covariance, gives $H=0.316$ for the periodic transform and $0.281$ for the mirror extension, so about half of the periodic bias is already present in the expectations, a departure of the finite-level energies from the asymptotic power law, and the remainder arises from taking logarithms of random energies. Normalization by the path energy does not contribute: it adds the same constant to every log share and leaves a least-squares slope with an intercept unchanged. Including the finest level increases the bias, to $-0.069$ for levels 3 to 8 and $-0.091$ for levels 4 to 8, so sampled fractional Brownian motion analyzed identically also places its finest level below the line; the low finest-level share of the wind record need not reflect sensor averaging. Calibrating the fit to exact finite-sample moments, as in the companion paper, accounts for the first effect exactly and for the second approximately.

The same measurement record, with a Hadamard gate on data qubit $0$, yields the within-level autocorrelations of Proposition~\ref{prop:lag}. Table~\ref{tab:wind} lists them together with the values implied by fractional Brownian motion with $H=1/3$, computed from the exact covariance of the path and the same circular db2 transform. The observed autocorrelations are negative at every level, with magnitudes near the model values at levels 4, 5, and 7. With $n=10^5$ preparations the readout standard deviation of $\hat r_j$ ranges from $0.008$ at level 3 to $0.029$ at level 8, so the departure at level 6 exceeds readout error and reflects the variability of a single path, which at that level contributes only 64 decimated coefficients per shift.

\begin{table}[htb]
\centering
{\footnotesize
\begin{tabular}{lcccccc}
\toprule
Level $j$ & 3 & 4 & 5 & 6 & 7 & 8\\
\midrule
Share $\bar p_j$ & 0.1363 & 0.0664 & 0.0476 & 0.0284 & 0.0156 & 0.0119\\
Autocorrelation $r_j$, data & $-0.341$ & $-0.247$ & $-0.200$ & $-0.396$ & $-0.188$ & $-0.144$\\
Autocorrelation $r_j$, fBm with $H=1/3$ & $-0.278$ & $-0.276$ & $-0.272$ & $-0.258$ & $-0.217$ & $-0.121$\\
\bottomrule
\end{tabular}
}
\caption{\small Quantum NDWT level shares and within-level lag-one autocorrelations (Proposition~\ref{prop:lag}) for the centered wind record, db2, $L=7$, with model values for fractional Brownian motion with $H=1/3$.}
\label{tab:wind}
\end{table}

\section{Resources, Scope, and Open Problems}
\label{sec:discussion}

The quantum NDWT acts on $J+L$ qubits and uses $O(pLJ)$ elementary gates for an orthonormal wavelet with $2p$ taps, of which $O(LJ)$ implement the controlled shifts and the remainder implement the single wavelet unitary. Reading level shares to relative precision $\delta$ costs $O\{(1-\bar p_j)/(\bar p_j\delta^2)\}$ preparations by sampling, somewhat fewer with the Fourier baseline, and $O\{1/(\delta\sqrt{\bar p_j})\}$ queries by amplitude estimation; within-level autocorrelations come from the same record at no additional cost. Cycle-spun linear shrinkage costs one run of the circuit of Fig.~\ref{fig:lcu}, which contains the quantum NDWT twice and a block encoding with $O(L^2)$ gates for level-wise weights or $O(LN)$ gates for translate-indexed weights, and succeeds with probability $\|\hat f\|^2/\|x\|^2$; learning the weights costs the fresh copies listed in Section~\ref{sec:cost}. These costs are polylogarithmic in $N$ except where noted, but they exclude the preparation of $\ket{y}$, which is $\Theta(N)$ for arbitrary classical data. The constructions therefore yield end-to-end savings only when the input state is produced coherently by an earlier quantum process or by an efficient structured loader, and the output, a quantum state, is consumed by further quantum processing or by estimation of a few functionals. Reconstructing all $N$ entries of $\hat f$ from copies of the heralded state requires on the order of $N/\eta^2$ preparations for accuracy $\eta$ and forfeits any advantage.

Within these limits, the baselines of Section~\ref{sec:measurement} and Proposition~\ref{prop:direct} determine what the shift-register construction contributes. Global shift-invariant quadratic statistics are estimated at least as well by a Fourier-basis measurement; every statistic of the data register alone, including the coefficient energies (\ref{eq:translate}), can be obtained with a classically randomized shift or from a unitary embedding of the nondecimated analysis; and the heralded shrinkage estimators are equally produced by direct analysis and synthesis or, for level-wise weights, by a Fourier-domain filter, with the same heralding probability. The distinctive feature of the construction is that a single circuit exposes the individual decimated branches and recombines them coherently. This supports branch-level operations, the comparison of decimated and nondecimated analyses within one record, amplitude estimation of level shares on the purified state with a one-gate marking oracle, and weighted recombination that interpolates between decimated and cycle-spun estimators. The unitary embedding of \citet{nakahira2021} uses $n+\lceil\log_2(L+1)\rceil$ qubits, 13 for $J=10$ and $L=5$ against our $J+L=15$ plus work qubits, and requires no postselection for the transform itself; the cost of its coefficient permutations is not analyzed there. Which route is cheapest for a given output map is a question of compiled gate counts, which we leave open.

Several problems remain open. The most important statistically is translation-invariant nonlinear thresholding with coherent access to the data, for which nonlinear amplitude transformations by quantum singular value transformation are the natural tool and for which the dependence of the cost on the sparsity of the wavelet representation is unknown. A second is data-driven selection of the constant and the window width in (\ref{eq:ew}), for instance by an unbiased risk estimate computed from the measurement record. A third is the extension to mirror-extended, multichannel, and two-dimensional transforms, where the shift register becomes a register of two-dimensional translations. Finally, the effect of hardware noise on the level shares, which enters through the depth of the controlled increments and the lattice delays, deserves a study on current devices.

\section*{Acknowledgments}
The author gratefully acknowledges support from the National Science Foundation under Grant No.~2515246 at Texas~A\&M~University and from the H.~O.~Hartley Chair research funds at Texas~A\&M. The author also thanks Nick~Broon of IBM~Quantum for valuable discussions during his visit to Texas~A\&M.

\bibliography{references03}

\end{document}